*Article*

# Impact of Surface Chemistry of Silicon Nanoparticles on the Structural and Electrochemical Properties of Si/Ni$_{3.4}$Sn$_4$ Composite Anode for Li-Ion Batteries

Tahar Azib [1], Claire Thaury [1,2], Fermin Cuevas [1,*], Eric Leroy [1], Christian Jordy [2], Nicolas Marx [3] and Michel Latroche [1]

[1] CNRS, ICMPE, Univ Paris Est Creteil, UMR 7182, 2 rue Henri Dunant, 94320 Thiais, France; azib@icmpe.cnrs.fr (T.A.); claire.thaury@gmail.com (C.T.); leroy@icmpe.cnrs.fr (E.L.); latroche@icmpe.cnrs.fr (M.L.)
[2] SAFT Batteries, 113 Bd. Alfred Daney, 33074 Bordeaux, France; Christian.JORDY@saftbatteries.com
[3] Umicore, Watertorenstraat 33, 2250 Olen, Belgium; Nicolas.Marx@eu.umicore.com
* Correspondence: cuevas@icmpe.cnrs.fr

**Abstract:** Embedding silicon nanoparticles in an intermetallic matrix is a promising strategy to produce remarkable bulk anode materials for lithium-ion (Li-ion) batteries with low potential, high electrochemical capacity and good cycling stability. These composite materials can be synthetized at a large scale using mechanical milling. However, for Si-Ni$_3$Sn$_4$ composites, milling also induces a chemical reaction between the two components leading to the formation of free Sn and NiSi$_2$, which is detrimental to the performance of the electrode. To prevent this reaction, a modification of the surface chemistry of the silicon has been undertaken. Si nanoparticles coated with a surface layer of either carbon or oxide were used instead of pure silicon. The influence of the coating on the composition, (micro)structure and electrochemical properties of Si-Ni$_3$Sn$_4$ composites is studied and compared with that of pure Si. Si coating strongly reduces the reaction between Si and Ni$_3$Sn$_4$ during milling. Moreover, contrary to pure silicon, Si-coated composites have a plate-like morphology in which the surface-modified silicon particles are surrounded by a nanostructured, Ni$_3$Sn$_4$-based matrix leading to smooth potential profiles during electrochemical cycling. The chemical homogeneity of the matrix is more uniform for carbon-coated than for oxygen-coated silicon. As a consequence, different electrochemical behaviors are obtained depending on the surface chemistry, with better lithiation properties for the carbon-covered silicon able to deliver over 500 mAh/g for at least 400 cycles.

## 1. Introduction

The rapid development of portable electronics, Electric Vehicles (EVs) and renewable energies requires light, safe and high-capacity rechargeable energy storage devices such as lithium-ion (Li-ion) batteries, one of the most efficient electrochemical storage systems today [1]. However, Li-ion batteries still need to be improved regarding design, electrode capacities and electrolyte stability [2]. Carbon-based anode materials are cheap and easy to prepare but suffer from moderate capacity (372 mAhg$^{-1}$ for graphite), which remains a limitation for the development of high-energy density storage [3]. Moreover, graphite suffers from parasitic reaction with the liquid electrolyte during charging and discharging processes to form the so-called Solid Electrolyte Interface (SEI), growth of which is detrimental for the stability and the capacity of the battery [4]. Therefore, new anode materials are required for the development of high-capacity Li-ion batteries.

Three main types of anode materials are currently envisaged for the replacement of graphite. Firstly, there are novel carbonaceous-based materials such as carbon nanotubes, carbon nanospheres, graphene and porous graphitic carbon [5–8]. Secondly, there are conversion-type transition metal compounds such as transition metal oxides, sulphides, selenides, fluorides, nitrides, phosphides and hydrides [9,10], and finally, there are silicon and tin-based anodes [11,12].

Pure *p*-type elements like Si or Sn are considered as promising to develop negative electrodes for Li-ion batteries [12–14]. Indeed, they can both be lithiated [15] to form binary compounds ($Li_{4.4}Sn$ and $Li_{3.75}Si$) with very large electrochemical capacities (994 and 3600 mAhg$^{-1}$, respectively) [12,16,17]. In addition to their high theoretical capacity, these elements have low potential and environmental friendliness. However, Si electrodes suffer from severe volume expansion during lithiation (up to 400%) [18]. Such swelling induces several drawbacks from the very first cycles like amorphization, delamination and capacity degradation, which are unfavorable for long term cycling [19–21].

To overcome these drawbacks, embedding the capacitive elements in a metallic matrix able to provide good electronic conductivity and to hold the volume changes is a beneficial solution [22]. This can be done with binary compounds having one element reacting with Li when the other one remains inactive, like for $Ni_3Sn_4$ [23–25], $Cu_6Sn_5$ [26], $CoSn_2$ [27], $FeSn_2$ [12], $NiSi_2$ [28] or $TiSi_2$ [29].

Following this concept, our group thoroughly investigated a composite of general formulation, Si-$Ni_{3.4}Sn_4$-Al-C, prepared via mechanochemistry [30–32]. It consists of submicronic silicon particles embedded in a nanostructured matrix made of $Ni_{3.4}Sn_4$, aluminum and graphite carbon. As reported by [33], low aluminum content (~3 wt.%) improves the cycle life of Si-Sn-type anodes. Carbon addition acts as a Process Control Agent (PCA), minimizing reactivity between Si and $Ni_{3.4}Sn_4$ on milling [30]. This composite takes advantage of the high capacities of silicon and tin, the good ionic and electronic conductivity of the matrix and its elastic properties to manage volume expansion. Further improvement for this composite can be foreseen by playing with the surface chemistry of the silicon particles [34,35].

In the present work, we investigate an alternative approach to PCA addition on milling that consists of modifying the surface chemistry of the silicon particles used for the composite synthesis. The Si surface is covered either with a carbonaceous or an oxide layer. Structural, morphological and electrochemical properties of these surface-modified silicon composites have been fully characterized and compared to those of non-modified Si. These new composites have very different properties, giving the best electrochemical performances for the carbon-coated silicon.

## 2. Materials and Methods

Three composites of Si-$Ni_{3.4}Sn_4$-Al were prepared using mechanochemistry of intermetallic $Ni_{3.4}Sn_4$ (75 wt.%), Al (3 wt.%) and three different kinds of silicon (22 wt.%): bare silicon, carbon- and oxide-coated silicon. Bare Si was provided by SAFT (purity 99.9%) as a reference for this work and is hereafter labelled as $Si_R$. The second Si precursor, labelled as $Si_C$, was provided by Umicore. Silicon particles were coated with carbon via Chemical Vapor Deposition (CVD) at 800 °C for 3 h. The third one ($Si_O$) was purchased from MTI Corporation as pure silicon. However, chemical analysis revealed that the particles were covered by a thin oxide layer. They were thus fully characterized regarding their surface chemistry and used as a Si-surface oxidized precursor ($Si_O$). These three Si-precursors were used to synthetize the composites (Si-$Ni_{3.4}Sn_4$-Al) via ball milling of Si, $Ni_{3.4}Sn_4$ (99.9%, ≤125 μm, home-made) and Al (99%, ≤75 μm, Aldrich) powders for 20 h under an inert atmosphere. Further details on intermetallic $Ni_{3.4}Sn_4$ and composite synthesis can be found in [30–32]. No addition of carbon graphite in the milling jar as PCA was used for the current investigation. The obtained composites are labelled as $Si_R$-NiSn, $Si_C$-NiSn and $Si_O$-NiSn, respectively.

X-Ray Diffraction (XRD) analysis of Si powders and composite materials was done with a Bruker D8 θ-θ diffractometer using Cu-K$\alpha$ radiation, in a 2θ range from 20 to 100° with a step size of 0.02°. Diffraction patterns were analyzed using the Rietveld method using the FULLPROF package [36]. Morphology of the composites was studied using Scanning Electron Microscopy (SEM) using a SEM-FEG MERLIN from Zeiss. Images were acquired from either Secondary Electrons (SE) or Back-Scattered Electrons (BSE) to provide information on particle morphology as well as phase distribution. Microstructural and chemical properties were analyzed using Transmission Electron Microscopy (TEM) with a Tecnai FEI F20 ST microscope providing high spatial resolution imaging of the scale morphology as well as chemistry via Energy-Dispersive X-ray spectroscopy (EDX) analyses. Images were taken in both bright and dark fields. Elemental mapping analysis was carried out using EDX analysis in scanning (STEM) mode and via Electron Energy Filtered Transmission Electron Microscopy (EFTEM). The samples were prepared by mixing the composite with Cu powder, followed by cold-rolling and thinning with argon ions in a GATAN precision ion polishing system.

Electrochemical measurements were carried out via galvanostatic cycling in half-coin type cells. A working electrode was prepared by mixing 40 wt.% of the 20-h-milled composite sieved under 36 μm, 30 wt.% of carboxymethyl-cellulose (CMC) binder and 30 wt.% of carbon black. Low loading of active material was adopted to avoid limitations on electrochemical performance due to electrode formulation. Metallic lithium was used as counter negative electrode separated by a 1 M solution electrolyte of $LiPF_6$ dissolved in Ethylene Carbonate (EC)/Propylene Carbonate (PC)/Dimethyl Carbonate (DMC) (1:1:3 *v/v/v*), supported by a microporous polyolefin Celgard™ membrane and a nonwoven polyolefin separator. The EC/PC/DMC mixture of carbonate-based solvent was selected based on its outstanding physico-chemical properties [37]. The battery was assembled in an argon filled glove box. The experiments were performed using a Biologic potentiostat instrument. To ensure full electrode lithiation, cells were cycled at C/50 for the first cycle, with a voltage window comprised between 0 and 2 V vs. Li+/Li, and at C/20 for the second and third cycles, with a voltage window comprised between 70 mV and 2 V vs. Li+/Li. The cut-off voltage of 70 mV was imposed to avoid the formation of crystalline $Li_{15}Si_4$ phase [38]. For all subsequent cycles, the kinetic regime was increased to C/10 to accomplish long-term cycling studies (up to 400 cycles) in a reasonable time duration and with a voltage window comprised between 70 mV and 2 V. Reference cycles at a rate of *C*/20 were done at second and third cycles and after every 20 cycles. Only the first and reference cycles are reported in this paper.

## 3. Results

### *3.1. Chemical and Microstructural Characterization*

3.1.1. Characterization of Bare and Surface-Modified Si Nanopowders

The XRD pattern of the Si$_R$ sample is displayed in Figure S1a (Supplementary Materials (SM)). All peaks can be indexed in the cubic space group *Fd-3m* with lattice constant $a = 5.426 \pm 2$ Å, slightly smaller than the well-crystallized silicon standard ($a = 5.430$ Å [39]). The measured crystallite size deduced from the diffraction peak linewidths is $16 \pm 2$ nm. SEM images reveal that the Si$_R$ powder has an interconnected worm-like morphology (Figure S1b). When observed using TEM, round particles with an average size of 180 nm are observed (Figure S1c). EFTEM analysis shows pure Si material with minor traces of oxygen at the surface (Figure S1d).

The Si$_C$ particles were chemically analyzed and contained Si (68 wt.%), C (30 wt.%) and O (2 wt.%). The Rietveld analysis of XRD patterns of Si$_C$ is shown in Figure S2a. The main phase is silicon ($a = 5.429 \pm 2$ Å, space group *Fd-3m*). The crystallite size is $79 \pm 3$ nm. A small and broad peak around 25°-2θ is attributed to the presence of poorly crystallized graphite (Figure S2a). SEM analysis reveals that the Si$_C$ powder is made of large agglomerates (10 to 100 μm) of primary spherical particles (Figure S2b). The particles were further investigated via TEM, confirming their spherical morphology (Figure S2c). In addition, TEM images show that the particles have an average size of 50 nm and are covered by a thin layer measuring a few nanometers (~10 nm) (Figure S2d). Elemental mapping indicates that the core of the particle is made of silicon surrounded by a thin carbon shell. At the interface, a composition gradient exists revealing a possible formation of silicon carbide SiC, assuring the chemical bonding between the two elemental layers.

The $Si_O$ particles were analyzed using XRD, showing that the main phase is silicon (cubic phase; $a$ = 5.426 ± 2 Å; space group *Fd-3m*) with a crystallite size around 26 ± 1 nm (Figure S3a). SEM analysis shows that the powder is made of large agglomerates up to 50 µm formed by primary submicrometric rounded particles (Figure S3b). From TEM analysis, it is observed that the primary particles are spherical with an average size of 70 nm (Figure S3c). Elemental analysis indicates a core of silicon surrounded by a shell containing both oxygen and silicon, which is attributed to the formation of $SiO_2$ (Figure S3d). A rough estimation based of the relative sizes of the core (58 nm in diameter) and the shell (6 nm thick) as well as the crystal densities of Si (2.33 g/cm$^3$) and $SiO_2$ (2.65 g/cm$^3$) leads to a global oxygen content in the $Si_O$ sample of 25 wt.%.

To summarize, Figure 1 displays TEM elemental mapping for the three types of Si nanopowders used in this study: bare Si, showing minor traces of oxygen at the surface, carbon-coated Si with a 10-nm-thick carbonaceous layer and oxide-coated Si with a 6-nm-thick $SiO_2$ oxide shell.

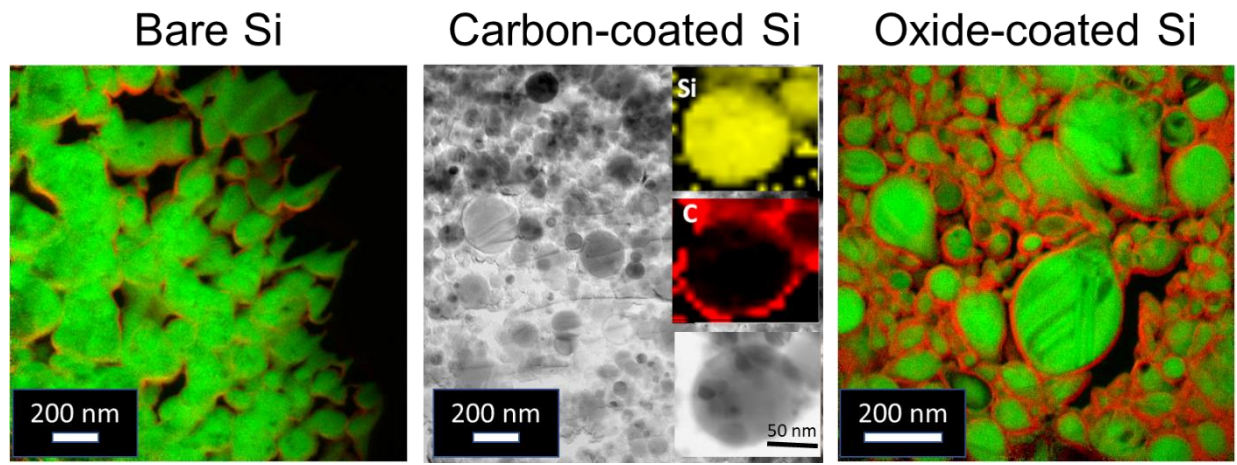

**Figure 1.** Transmission Electron Microscopy (TEM) images and elemental mapping of Si nanoparticles used as precursors for composite synthesis. (**Left**) bare Si, $Si_R$ (Si in green, oxygen traces in red), (**center**) carbon-coated Si, $Si_C$ (Si in yellow, carbon in red) and (**right**) oxide-coated Si, $Si_O$ (Si in green, oxygen in red).

3.1.2. Characterization of the Composite Materials

Three composites, $Si_R$-NiSn, $Si_C$-NiSn and $Si_O$-NiSn, were synthetized using the previously analyzed silicon nanopowders. The weight and atomic compositions of the composites are given in Table 1. The amount of carbon and oxygen for each composite are estimated from the chemical analysis of the surface-modified Si particles assuming a shell of pure C for $Si_C$ and a shell of $SiO_2$ for $Si_O$.

**Table 1.** Compositions of $Si_R$-NiSn, $Si_C$-NiSn and $Si_O$-NiSn composites.

| Composites | Weight Composition | Atomic Composition |
|---|---|---|
| $Si_R$-NiSn | $Si_{0.22}Ni_{0.22}Sn_{0.53}Al_{0.03}$ | $Si_{0.46}Ni_{0.22}Sn_{0.26}Al_{0.06}$ |
| $Si_C$-NiSn | $Si_{0.15}Ni_{0.22}Sn_{0.53}Al_{0.03}C_{0.07}$ | $Si_{0.26}Ni_{0.18}Sn_{0.22}Al_{0.06}C_{0.28}$ |
| $Si_O$-NiSn | $Si_{0.17}Ni_{0.22}Sn_{0.53}Al_{0.03}O_{0.06}$ | $Si_{0.32}Ni_{0.20}Sn_{0.24}Al_{0.05}O_{0.19}$ |

Evolution of the diffractograms for the three composites as a function of milling time between 1 and 20 h is shown in Figure S4. For Si$_R$-NiSn, diffraction peaks of the intermetallic precursor Ni$_{3.4}$Sn$_4$ progressively disappear, Si peaks broaden and new peaks due to Sn formation appear. For Si$_C$-NiSn, diffraction peaks of Ni$_{3.4}$Sn$_4$ and Si are preserved though undergoing significant line broadening. A minor contribution of Sn formation is detected. Finally, for Si$_O$-NiSn, Ni$_{3.4}$Sn$_4$ and Si are mostly preserved, but compared to the Si$_C$-NiSn composite, peak broadening for the intermetallic precursor is less pronounced and a secondary intermetallic phase of Ni$_3$Sn$_4$ with lower Ni-content than that of the pristine precursor Ni$_{3.4}$Sn$_4$ is formed. A minor contribution of Sn formation is also detected.

The XRD diffraction patterns for the 20-h-milled composites are displayed in Figure 2. Rietveld analysis is provided in Figure S5 and collected crystal data are gathered in Table 2. For the composite made with bare Si, Si$_R$-NiSn, major phases are Sn (43 ± 1 wt.%) and NiSi$_2$ (35 ± 1 wt.%). These phases result from a mechanically-induced chemical reaction between the milling precursors Ni$_{3.4}$Sn$_4$ and Si. In contrast, for the coated composites Si$_C$-NiSn and Si$_O$-NiSn, the main phase remains Ni$_{3+x}$Sn$_4$-type (~85 wt.%) evidencing minor chemical reaction between Ni$_{3.4}$Sn$_4$ and Si on milling. Indeed, after 20 h of milling, the content of Sn byproduct in the Si-coated composites is as low as ~3 wt.%. Nonetheless, it is worth noticing that for the Si$_O$-NiSn composite, almost half of the pristine intermetallic precursor Ni$_{3.4}$Sn$_4$ (34 ± 2 wt.%) diminishes in Ni-content to form Ni$_3$Sn$_4$. In addition, note that the crystallite size for Ni$_{3.4}$Sn$_4$ is much smaller for Si$_C$-NiSn (7 ± 2 nm) than for Si$_O$-NiSn (39 ± 3 nm).

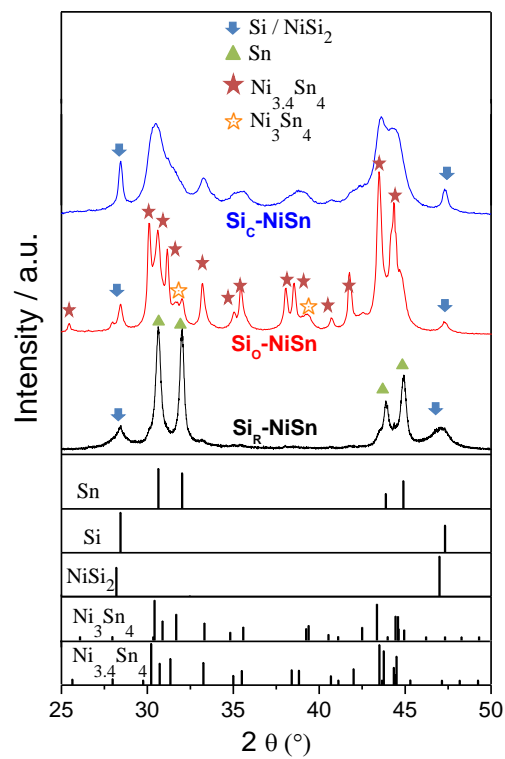

**Figure 2.** X-Ray Diffraction **(**XRD) patterns of the composites Si$_R$-NiSn, Si$_C$-NiSn and Si$_O$-NiSn after 20 h of milling. Position of diffraction lines for Sn, Si, NiSi$_2$, Ni$_3$Sn$_4$ and Ni$_{3.4}$Sn$_4$ phases as reported in Pearson's crystal data base [40] are shown in the bottom part of the figure.

**Table 2.** Crystallographic data for the Si$_R$-NiSn, Si$_C$-NiSn and Si$_O$-NiSn composites after 20 h of milling as determined from Rietveld analysis. Ni over-stoichiometry (x) in Ni$_{3+x}$Sn$_4$ and crystallite size (L in nm) for all phases are given. Standard deviations refereed to the last digit are given in parenthesis.

| Sample | Phases | Content (wt.%) | S.G. | Cell Parameters | | | | x in Ni$_{3+x}$Sn$_4$ | L (nm) | $R_B$ | $R_{wp}$ |
| --- | --- | --- | --- | --- | --- | --- | --- | --- | --- | --- | --- |
| | | | | a (Å) | b (Å) | c (Å) | β(°) | | | | |
| Si$_R$-NiSn | Ni$_3$Sn$_4$ | 9(1) | C2/m | 12.199 * | 4.0609 * | 5.2238 * | 105.17 * | 0* | 10 * | 6.8 | 9.7 |
| | Si | 13(1) | Fd$\bar{3}$m | 5.430 * | | | | | 15(2) | 2.5 | |
| | Sn | 43(1) | I4$_1$/amd | 5.8303(2) | | 3.1822(1) | | | 27(1) | 2.7 | |
| | NiSi$_2$ | 35(2) | Fm$\bar{3}$m | 5.4731 (5) | | | | | 5(1) | 3.6 | |
| Si$_C$-NiSn | Ni$_{3.4}$Sn$_4$ | 85(2) | C2/m | 12.357 (3) | 4.060(1) | 5.201(2) | 104.31(2) | 0.45(9) | 7(2) | 2.4 | 4.8 |
| | Si | 12(1) | Fd$\bar{3}$m | 5.431(1) | | | | | 30(2) | 5.2 | |
| | Sn | 2(1) | I4$_1$/amd | 5.8303 * | | 3.1822 * | | | 27 * | 4.5 | |
| | NiSi$_2$ | 1(1) | Fm$\bar{3}$m | 5.4731 * | | | | | 5 * | 2.4 | |
| Si$_O$-NiSn | Ni$_{3.4}$Sn$_4$ | 49(2) | C2/m | 12.448(2) | 4.079(1) | 5.209(1) | 103.62(1) | 0.4* | 39(3) | 7.1 | 7.8 |
| | Ni$_3$Sn$_4$ | 34(2) | C2/m | 12.248(3) | 4.046 (1) | 5.201(1) | 104.88(1) | 0* | 14(2) | 5.9 | |
| | Si | 12(2) | Fd$\bar{3}$m | 5.433(2) | | | | | 19(2) | 13.3 | |
| | Sn | 4(1) | I4$_1$/amd | 5.8303 * | | 3.1822 * | | | 27 * | 4.7 | |
| | NiSi$_2$ | 1(1) | Fm$\bar{3}$m | 5.4731 * | | | | | 5 * | 17.3 | |

\* fixed values.

The morphology of the three composites after 20 h of milling was examined using SEM and is displayed in Figure 3. The composite Si$_R$-NiSn consists of micrometric-size round-shaped particles (Figure 3a). The composite particles contain phase domains of dark tonality attributed to silicon nanoparticles [30] embedded in a light-grey matrix which is chemically homogeneous at the spatial resolution (~50 nm) of the BSE analysis (Figure 3b). In the case of material ground with Si$_C$, SEM-SE analysis (Figure 3c) shows that the composite particles are in the form of micrometer-sized platelets. SEM-BSE analysis (Figure 3d) reveals that the platelets are formed by particles with dark tonality (attributed to silicon) surrounded, as for the previous composite, by a chemically homogeneous light-grey matrix. Note that the silicon particle size (dark domains) is comparable for Si$_R$-NiSn and Si$_C$-NiSn composites. There are also brighter areas attributed to some Ni$_{3+x}$Sn$_4$ domains of micrometric size. Figure 3e,f show the SEM images for Si$_O$-NiSn composite. The composite particles also form platelets in the micrometric range. In the BSE-SEM image (Figure 3f), it is observed that the phase distribution within the particles is very inhomogeneous. There are very dark areas attributed to agglomerates of silicon particles and other areas with two different grayscales ascribed to the intermetallic Ni$_{3+x}$Sn$_4$ phase.

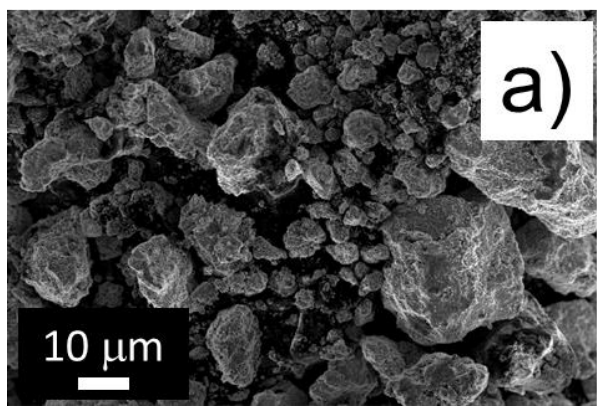
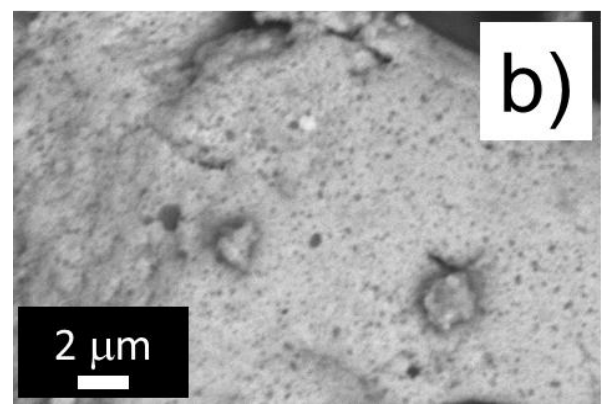

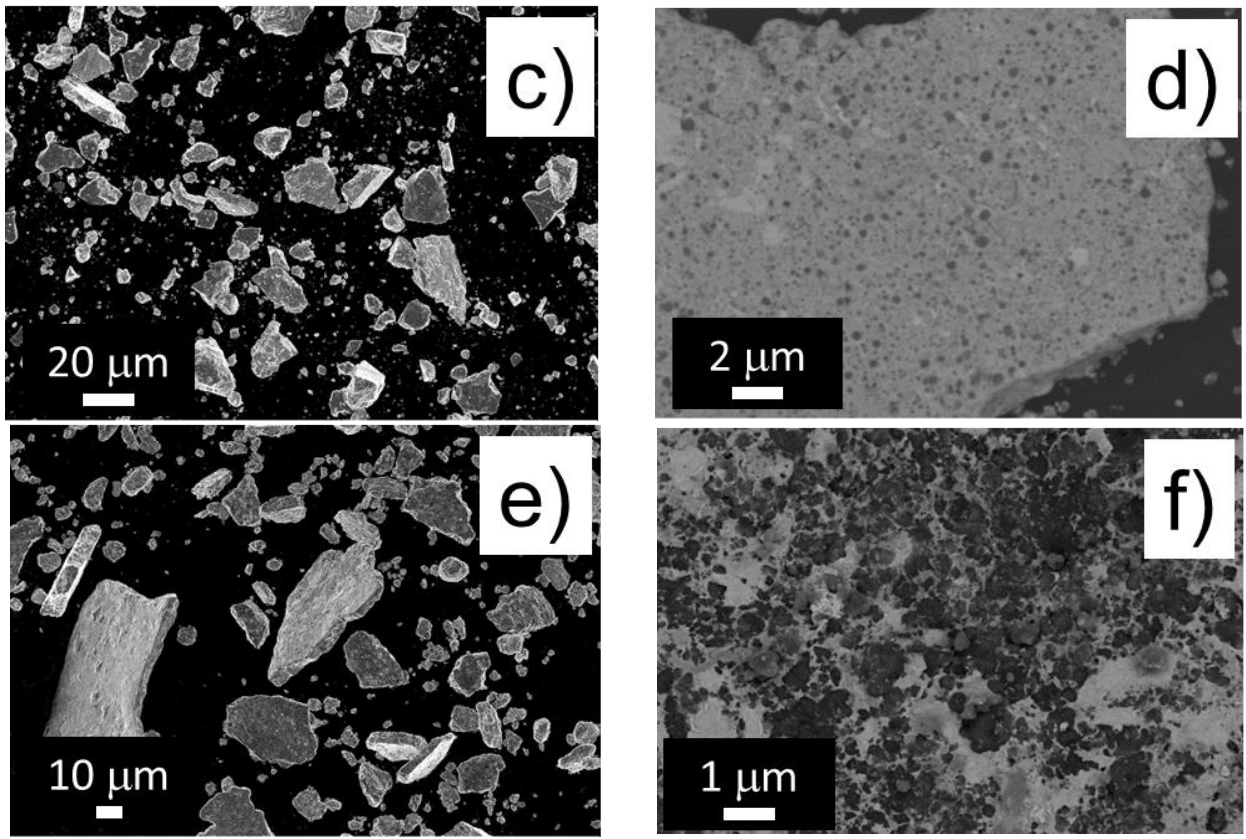

**Figure 3.** Scanning Electron Microscopy (SEM) images for Si$_R$-NiSn (**a**,**b**) Si$_C$-NiSn (**c**,**d**) and Si$_O$-NiSn (**e**,**f**) composites. Images were taken in either Secondary Electrons (SE) (**a**,**c**,**e**) or Back-Scattered Electrons (BSE) (**b**,**d**,**f**) modes.

To get a more accurate analysis of the chemically-homogeneous matrix in Si$_R$-NiSn and Si$_C$-NiSn composites, TEM analyses were performed (Figure 4). For the composite Si$_R$-NiSn (Figure 4, top), Si nanoparticles are surrounded by all elements. There is not a complete spatial correlation between Ni and Sn signals, which corroborates the decomposition of Ni$_{3.4}$Sn$_4$ as observed using XRD (Figure 2), leading to the formation of free Sn at the nanoscale. The analysis of the Si$_C$-NiSn composite (Figure 4, bottom) shows that the silicon particles are surrounded by a homogeneous matrix that contains Ni, Sn, C and Al. The Ni and Sn signals are spatially correlated indicating the presence of the Ni-Sn intermetallic at the nanometer scale in agreement with XRD results (Figure 2, Table 2). No preferential distribution of carbon is seen around the silicon particles: the carbon layer may have been dissolved upon grinding. However, carbon mapping should be considered with caution as carbon deposition is likely to occur under the electron beam. Complementary high-resolution TEM analysis (Figure S5) confirms the size of the coherent domains calculated by Rietveld refinement for the two main phases: about 30 nm for silicon (red area) and 8 nm for the Ni-Sn phase (black area).

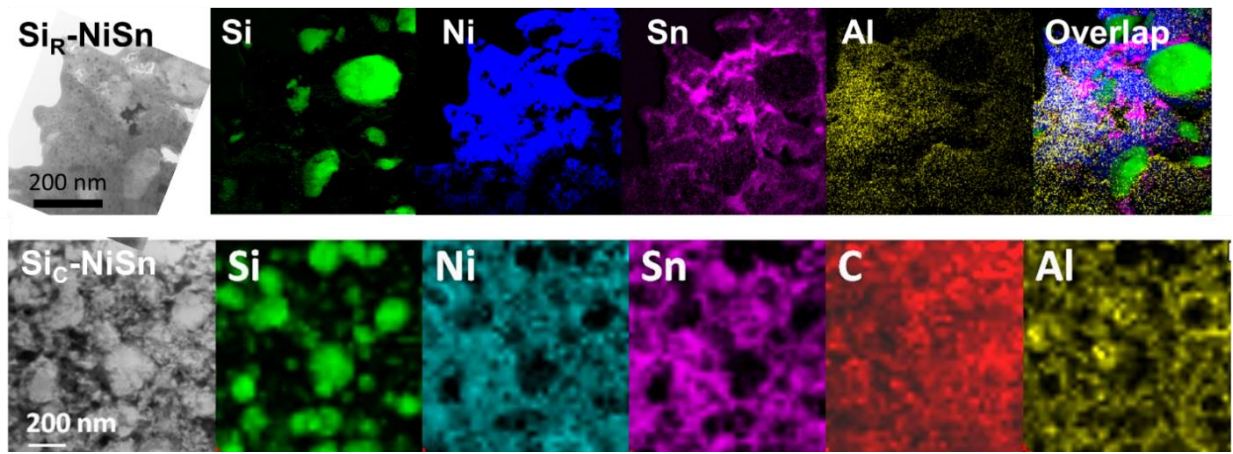

**Figure 4.** TEM images and elemental mapping for Si$_R$-NiSn ((**top**), Electron Energy Filtered Transmission Electron Microscopy (EFTEM) mapping) Si$_C$-NiSn ((**bottom**), STEM-EDX mapping).

### 3.2. Electrochemical Characterization

Figure 5 displays the potential profiles of the three studied Si-NiSn composites for the first and third cycles. At the first cycle, discharge (lithiation) profiles show a shoulder at 1.25 V attributed to the formation of the Solid Electrolyte Interface (SEI). Then, the potential profiles gradually decrease down to 0 V, showing several steps (~0.65, 0.40 and 0.35 V) for Si$_R$-NiSn, while the coated Si$_C$-NiSn and Si$_O$-NiSn composites have smooth potential profiles. Among the three composites, Si$_C$-NiSn has the lowest polarization potential. The first lithiation capacity is much lower for the oxide-coated Si$_O$-NiSn (685 mAh/g) than for Si$_R$-NiSn and Si$_C$-NiSn (950 and 1195 mAh/g, respectively). On charge (delithiation), either smooth or staircase potential profiles are again observed for coated (Si$_C$-NiSn and Si$_O$-NiSn) and bare (Si$_R$-NiSn) composites, respectively. At the third cycle, potential profiles show no evidence of SEI formation, and similarly to the 1st cycle, they are smooth for the coated composites while the bare composite has a staircase potential profile.

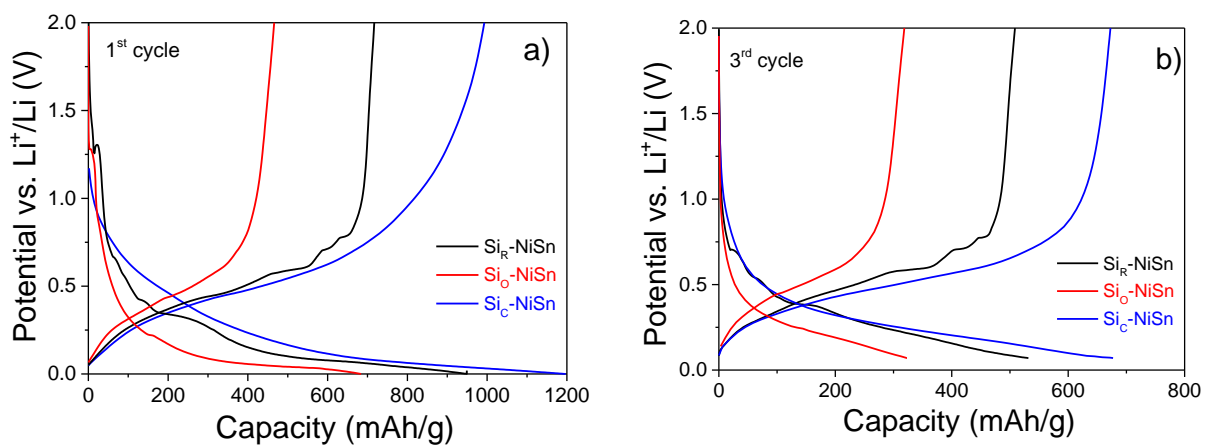

**Figure 5.** Discharge/charge profiles of Si$_R$-NiSn, Si$_O$-NiSn and Si$_C$-NiSn composites at the first (**a**) and third (**b**) galvanostatic cycle.

The evolution of reversible capacities (delithiation) and coulombic efficiency on cycling for the three composites is shown in Figure 6. For all composites, a significant capacity decay is observed during the first three activation cycles. Then, for the bare Si$_R$-NiSn composite, the capacity gradually decreases from 400 mAh/g at cycle 25 down to 210 mAh/g at cycle 200. In contrast, for the coated composites the capacity remains stable on cycling after activation, being significantly higher for Si$_C$-NiSn than for Si$_O$-NiSn. After 400 cycles, their reversible capacities are 505 and 215 mAh/g, respectively. As for the coulombic efficiency, $\varepsilon_c$ (Figure 6b), it strongly depends on the composite at the first cycle. It ranges between 68% for oxide-coated Si$_O$-NiSn and 83% for the carbon-coated Si$_C$-NiSn composite. For the next cycles, the coulombic efficiencies drastically increase for all composites with typical values above 99.5%.

To summarize, from the three studied composites, Si$_C$-NiSn exhibits the best electrochemical properties with a reversible capacity exceeding 500 mAh/g over 400 cycles. It has a reasonable initial coulombic efficiency of 83%, which increases to an average value of 99.6% between reference cycles 3 to 400.

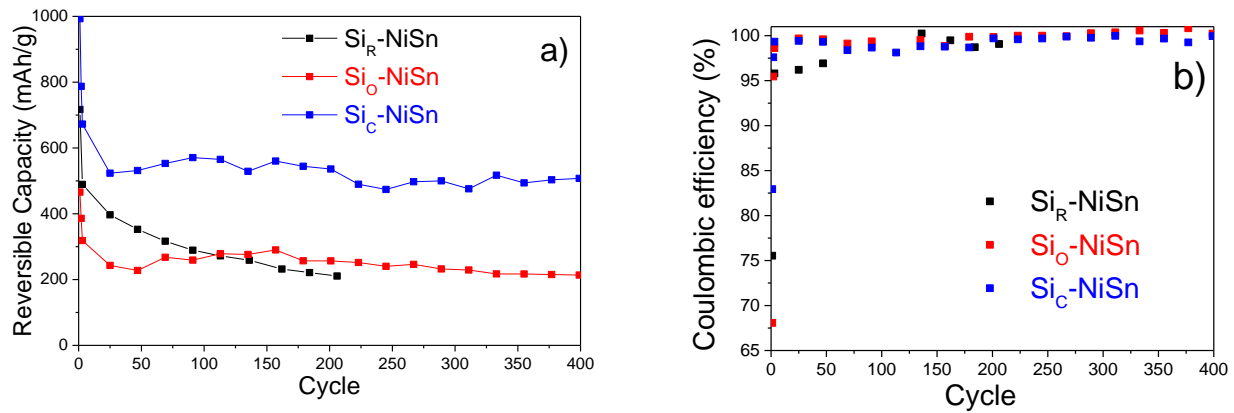

**Figure 6.** Evolution of the specific reversible capacity (**a**) and coulombic efficiency (**b**) of Si$_R$-NiSn, Si$_O$-NiSn and Si$_C$-NiSn composites during galvanostatic cycling.

## 4. Discussion

In this work, silicon nanoparticles with bare and chemically modified surfaces (with either C or O) have been used to prepare composites of Si-Ni$_{3.4}$Sn$_4$-Al using mechanical milling. Bare Si nanoparticles used as reference, Si$_R$, have an average size of 180 nm and contain minor traces of oxygen at the surface. The surface-modified Si-particles are nanometric (around 50–70 nm) and exhibit a core-shell structure with a shell thickness of 10 and 6 nm for Si$_C$ and Si$_O$, respectively. Chemical analyses show that, for Si$_C$, the shell is mainly composed of carbon representing 30 wt.%C, whereas the shell chemistry of Si$_O$ is identified as a silicon oxide with 46 wt.% of SiO$_2$, i.e., 25 wt.% of oxygen.

The chemical, microstructural and electrochemical properties of the Si-NiSn composites prepared with the different types of Si nanoparticles are summarized in Table 3. The ground composites consist of micrometer-sized particles that are round for bare silicon, Si$_R$, but turn to be platelet-like when using coated silicon Si$_O$ and Si$_C$. Differences in composite morphology are ascribed to the fact that the Si surface chemistry plays a major role in the phase composition upon milling. Indeed, for coated Si, the major phase is the intermetallic Ni$_{3+x}$Sn$_4$, while Sn is the major phase when using bare Si. Formation of tin on milling results from the reaction

$$6Si + Ni_3Sn_4 \xrightarrow{milling} 3NiSi_2 + 4Sn \tag{R1}$$

which produces NiSi$_2$ in addition as a secondary phase. The stoichiometry of the Ni$_{3+x}$Sn$_4$ phase is not considered here for the sake of simplicity. The occurrence of ductile Sn favors the formation of round-shaped particles on milling [41].

Table 3. Summary of the chemical, microstructural and electrochemical properties of the three composites Si$_R$-NiSn, Si$_C$-NiSn and Si$_O$-NiSn.

| Composites | Si$_R$-NiSn | Si$_C$-NiSn | Si$_O$-NiSn |
| --- | --- | --- | --- |
| Composition (wt.%) | Si$_{0.22}$Ni$_{0.22}$Sn$_{0.53}$Al$_{0.03}$ | Si$_{0.15}$Ni$_{0.22}$Sn$_{0.53}$Al$_{0.03}$C$_{0.07}$ | Si$_{0.17}$Ni$_{0.22}$Sn$_{0.53}$Al$_{0.03}$O$_{0.06}$ |
| Composition (at.%) | Si$_{0.46}$Ni$_{0.22}$Sn$_{0.26}$Al$_{0.06}$ | Si$_{0.26}$Ni$_{0.18}$Sn$_{0.22}$Al$_{0.06}$C$_{0.28}$ | Si$_{0.32}$Ni$_{0.20}$Sn$_{0.24}$Al$_{0.05}$O$_{0.19}$ |
| Main phase (XRD; wt.% ±x) | Sn (43 ± 1) | Ni$_{3+x}$Sn$_4$ (85 ± 2) | Ni$_{3+x}$Sn$_4$ (83 ± 3) |
| Matrix phase distribution | Homogeneous | Homogeneous | Heterogenous |
| Particles morphology | Round-shaped | Platelets | Platelets |
| Sn phase (XRD; wt.% ±x) | 43 ± 1 | 2 ± 1 | 4 ± 1 |
| Crystal size Ni$_{3+x}$Sn$_4$ (nm) | 9 | 7 | 14–39 |
| Crystal size Si (nm) | 15 ± 2 | 30 ± 2 | 19 ± 2 |
| Potential profiles | Staircase | Smooth | Smooth |
| C$_{rev}$ (1st cyc.; mAhg$^{-1}$) | 715 | 995 | 465 |
| C$_{rev}$ (3rd cyc.; mAhg$^{-1}$) | 490 | 675 | 320 |
| C$_{rev}$ (mAhg$^{-1}$)@cycle# | 210@cycle200 | 505@cycle400 | 215@cycle400 |
| ε$_C$ (1st cycle; %) | 75 | 83 | 68 |
| ε$_C$ (aver. 3–400 cycles; %) | - | 99.6 | 99.8 |

Clearly, Si coating minimizes reaction R1, preserving the initial reactants, Si and Ni$_{3.4}$Sn$_4$, by avoiding direct contact between the two phases on milling. However, reaction R1 is not fully suppressed on prolonged milling since 2 and 4 wt.% of Sn are detected using XRD for Si$_C$ and Si$_O$, respectively. This reveals that carbon coating is more efficient than the oxide one, which is further supported by the fact that the stoichiometry of Ni$_{3+x}$Sn$_4$ remains constant for the Si$_C$-NiSn composite while it is partially altered for Si$_O$-NiSn (Table 2) [38]. The lower efficiency of the oxide coating to minimize the reaction between Si and Ni$_{3.4}$Sn$_4$ can be tentatively attributed to its small thickness (6 nm) and to the fact that the coating also contains Si in the form of SiO$_2$. Interestingly, it should also be noted that carbon coating not only minimizes Sn formation but also enhances the nanostructuration of the Ni$_{3+x}$Sn$_4$ phase. The crystallite size of Ni$_{3+x}$Sn$_4$ is much smaller with the carbon coating (crystallite size $L \sim 7$ nm) than for the oxide one ($L \sim 14$–39 nm). Thus, carbon coating allows efficient nanostructuration of the matrix leading to good chemical homogeneity around the Si nanoparticles (Figure 3).

The difference in chemical and microstructural properties between Si$_C$-NiSn, Si$_O$-NiSn and Si$_R$-NiSn composites lead to clearly distinct electrochemical behaviors, which are also summarized in Table 3. The reference Si$_R$-NiSn composite has staircase potential profiles, moderate initial capacity and poor cycle-life. Oxide coating of Si nanoparticles leads to smooth potential profiles and good cycle-life but at the expense of limited capacity. Finally, carbon coating not only lead to smooth potential profiles but also to high capacity and coulombic efficiency over hundreds of cycles. Smooth profiles are preferred to staircase ones since the volume changes of the active materials induced by phase transformations occur gradually for the former, minimizing mechanical degradation on cycling. A better insight into the different electrochemical properties between the composites can be gained at the light of the microstructural properties and through deep analysis of potential profiles (Figure 5) by evaluating Differential Capacity Plots (DCPs).

Figure 7 displays the DCP plots for the three composites at the 1st, 3rd and 400 cycles. For the first galvanostatic cycle of the bare Si$_R$-NiSn composite (Figure 7a), four clear reduction peaks of moderate intensity are observed at 0.66, 0.56, 0.42 and 0.34 V and a broad additional one below 0.1 V. The first four peaks are attributed to lithiation of the main phase (free Sn), whereas the latter one is assigned to the formation of Li-rich Li$_y$Si and Li$_z$Sn alloys [31]. In the anodic branch, four clear oxidation peaks are observed at 0.44, 0.58, 0.70 and 0.78 V, which are attributed to the decomposition of the different Li$_z$Sn alloys (Li$_7$Sn$_2$, Li$_5$Sn$_2$, LiSn and Li$_2$Sn$_5$) in agreement with previous reports [38,42]. The signal at 0.58 V is in fact a triplet due to the decomposition of three Li$_z$Sn alloys of close composition: Li$_{13}$Sn$_5$, Li$_5$Sn$_2$ and Li$_7$Sn$_3$ [42,43]. In addition, an anodic bump and a broad oxidation peak can be observed at 0.32 and 0.46 V, which are attributed to the decomposition of amorphous $Li_{3.16}Si$ and $Li_7Si_3$ alloys, respectively [20,38,44]. The detected $dQ/dV$ peaks for the first galvanostatic cycle in the bare $Si_R$-NiSn composite are consistent with the coexistence of pure Si and Sn phases (Table 2). The width of the peaks is anticorrelated with the crystallinity of the phases: anodic peaks due to decomposition of amorphous $Li_ySi$ alloys formed during the first composite lithiation [45] are wider than those of the crystalline $Li_zSn$ ones [46]. Interestingly, at the third cycle (Figure 7b), $dQ/dV$ peaks attributed to the formation and decomposition of $Li_zSn$ alloys are sharper than those of the first cycle, which suggests the coarsening of Sn domains on cycling [47]. This agglomeration favorizes discrete volume changes, leading to electrode cracking [19] and severe capacity decay for the bare $Si_R$-NiSn composite, as observed in Figure 6a.

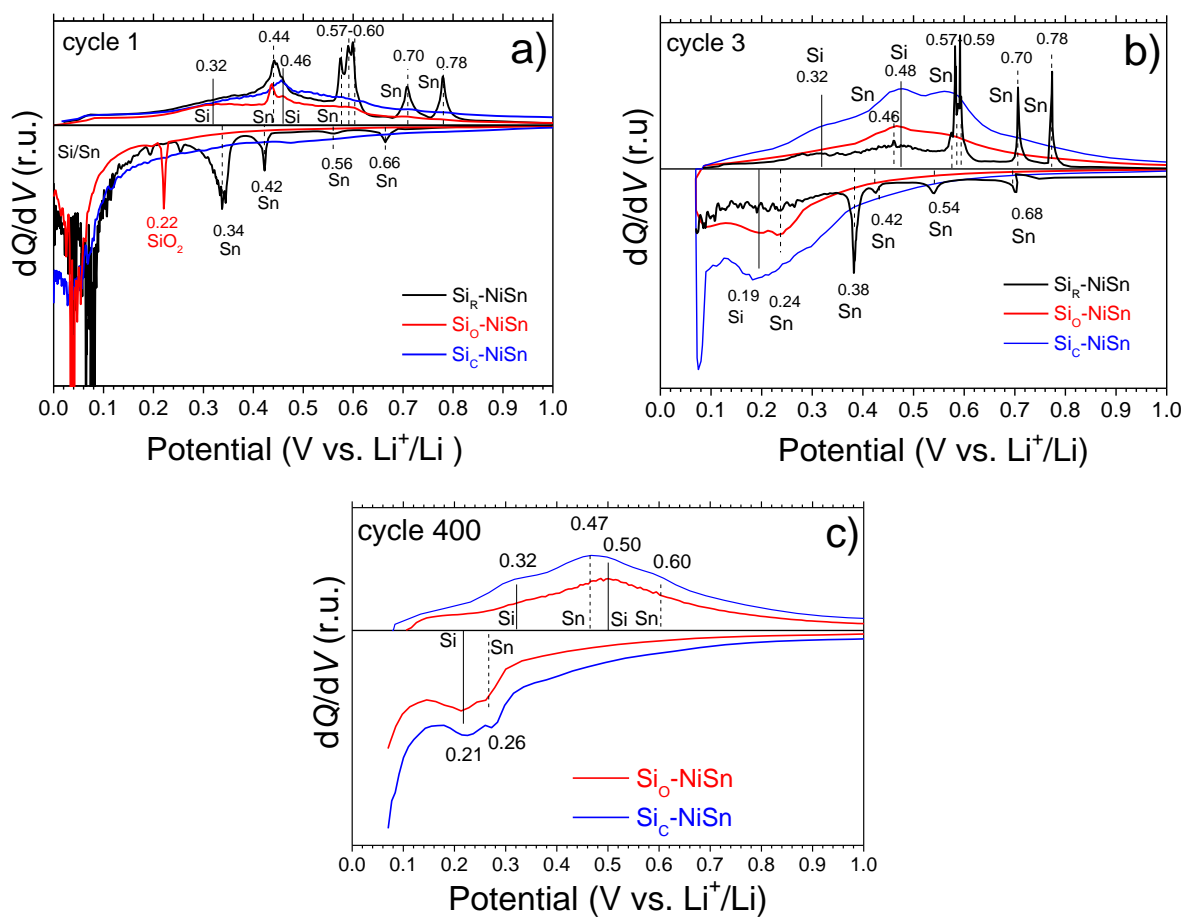

**Figure 7.** Differential capacity plots for all composites at cycles (**a**) 1, (**b**) 3 and (**c**) 400. Solid and dashed vertical lines have been used to identify formation/decomposition of $Li_ySi$ and $Li_zSn$ alloys, respectively.

The $dQ/dV$ plot for the oxide-coated $Si_O$-NiSn composite at the 1st cycle (Figure 7a) displays two reduction peaks at 0.22 and 0.04 V which are assigned to the lithiation of silicon oxide $SiO_2$ and crystalline Si, respectively [45,48]. The anodic branch at the 1st cycle is almost featureless with a broad peak at 0.46 V and a shoulder at ~0.59 V attributed to decomposition of $Li_zSn$ alloys, as well as two shoulders at 0.32 and 0.46 V assigned to decomposition of the $Li_ySi$ ones. The broadness of these signals is a signature of the low crystallinity of the reacting phases for this composite. This strongly differs from the well-defined oxidation peaks observed for $Si_R$-NiSn, which evidences a different chemical and microstructural state of Sn between the $Si_R$-NiSn and $Si_O$-NiSn composites. For $Si_R$-NiSn, free Sn is formed during ball milling (Table 2) with a crystallite size of 27 nm. As men-tioned before, free Sn likely coarsens due to agglomeration during electrochemical cycling. In contrast, for the Si$_O$-NiSn composite, Sn remains alloyed with Ni in the form of nanometric Ni$_3$Sn$_4$ intermetallic after ball milling. Upon lithiation/delithiation, Ni$_3$Sn$_4$ undergoes a reversible conversion reaction which can be expressed as [23,38,49]

$$Ni_3Sn_4 + 14Li \leftrightarrow 2Li_7Sn_2 + 3Ni \tag{R2}$$

This conversion reaction ensures the nanostructured state of Sn-forming alloys, Ni$_3$Sn$_4$ and Li$_z$Sn, on cycling [12]. It should be also noticed that very similar featureless anodic branches are observed at cycles 3 (Figure 7b) and 400 (Figure 7c), which is concomitant with the long-term cycling stability of Si$_O$-NiSn (Figure 6a). As for the cathodic branch at cycles 3 and 400, two broad peaks are detected at 0.24 and 0.19 V that are tentatively attributed to the formation of poorly crystallized Li$_z$Sn and Li$_y$Si alloys, respectively [31]. The sharp peak detected at 0.22 V at the 1st reduction attributed to the SiO$_2$ lithiation is not detected in subsequent cycles showing its irreversible behavior. Indeed, as reported by Guo et al. [48], SiO$_2$ can react with lithium through two reaction paths:

$$SiO_2 + 4Li^+ + 4e^- \rightarrow 2Li_2O + Si \tag{R3}$$

$$2SiO_2 + 4Li^+ + 4e^- \rightarrow Li_4SiO_4 + Si \tag{R4}$$

The irreversibility of these reactions accounts for the low coulombic efficiency of Si$_O$-NiSn in the first cycle (68%, Figure 6b) [50] and explains the low reversible capacity of this composite. In addition, the effect of the poor chemical homogeneity of the composite matrix (Figure 3f) on the limited first lithiation capacity (685 mAh/g) of the Si$_O$-NiSn composite cannot be ruled out (Figure 5a).

Finally, the dQ/dV plots for the carbon-coated Si$_C$-NiSn composite exhibit, as a general trend, smooth traces both for cathodic and anodic branches and all over the 400 cycles (Figure 7). At the first reduction, a unique clear peak below 0.1 V is detected and attributed to formation of Li-rich Li$_y$Si alloys and the conversion reaction R2 for the major Ni$_3$Sn$_4$ phase. No signal of large SEI formation is observed, which concurs with the high initial coulombic efficiency of this composite (87%). At the 1st oxidation, a broad peak at 0.46 V is attributed to decomposition of Li$_y$Si alloys, while bumps at ~0.44 and 0.58 V can be assigned to decomposition of Li$_z$Sn alloys leading to the recovery of Ni$_3$Sn$_4$ [31]. At cycles 3 and 400, very similar and smooth curves are detected showing several bumps that point out good and stable reversibility in the lithiation of Si and Ni$_3$Sn$_4$ counterparts of the Si$_C$-NiSn composite. It should be noted that the area under the dQ/dV plots is much larger for Si$_C$-NiSn than for Si$_O$-NiSn showing the higher capacity of the former (Figure 6).

## 5. Conclusions

Modification of the Si surface chemistry clearly affects the chemical and microstructural properties of Si-Ni$_{3.4}$Sn$_4$-Al composites as anode materials in Li-ion cells. First, it plays a protective role in the mechanochemical synthesis of the composite. This is indeed an effective solution for limiting the reaction between silicon and Ni$_{3.4}$Sn$_4$ during grinding and thus preventing the formation of detrimental free Sn. The milling process with Si-coated particles leads to a platelet-like morphology of the composites for both oxide- and carbon-coated silicon in contrast with the round-shaped one using bare silicon. However, differences in the microstructure of the composite matrix are found as a function of the surface chemistry, being chemically heterogeneous at the nanoscale for oxide coating while it is homogeneous for the carbon one. This leads to a low lithiation capacity for oxide coating and, moreover, low coulombic efficiency at the first cycle due to an irreversible reaction between SiO$_2$ and lithium. The use of carbon coating leads to a homogeneous matrix surrounding Si nanoparticles leading to a high reversible capacity that keeps stable after hundreds of cycles. Such an approach allows high performance materials usable as anodes for high energy density batteries to be developed.

**Supplementary Materials:** The following are available online at www.mdpi.com/xxx/s1, Figure S1: Microstructural characterization of bare Si powder, Figure S2: Microstructural characterization of carbon-coated Si powder, Figure S3: Microstructural characterization of oxide-coated Si powder, Figure S4: Evolution of the XRD patterns of Si-NiSn composites as a function of milling time, Figure S5: Graphical output of Rietveld analysis of Si-NiSn composites, Figure S6: High-resolution TEM image of the Si$_c$ -NiSn composite.

**Author Contributions:** Conceptualization, T.A., C.T., F.C., C.J., N.M., M.L.; methodology, C.T., F.C., E.L., M.L.; validation, F.C., C.J., N.M., M.L.; investigation, T.A., C.T., F.C., M.L.; writing—original draft preparation, T.A., C.T., F.C., M.L.; writing—review and editing, T.A., C.T., F.C., E.L., C.J., N.M., M.L.; supervision, F.C., M.L.; project administration, F.C.; funding acquisition, F.C., M.L., C.J., N.M. All authors have read and agreed to the published version of the manuscript.

**Funding:** This research was funded by the the French Research Agency (ANR), project NEW-MASTE, grant number n° ANR-13-PRGE-0010.

**Institutional Review Board Statement:** In this section, you should add the Institutional Review Board Statement and approval number, if relevant to your study. You might choose to exclude this statement if the study did not require ethical approval. Please note that the Editorial Office might ask you for further information. Please add "The study was conducted according to the guidelines of the Declaration of Helsinki, and approved by the Institutional Review Board (or Ethics Committee) of NAME OF INSTITUTE (protocol code XXX and date of approval)." OR "Ethical review and approval were waived for this study, due to REASON (please provide a detailed justification)." OR "Not applicable" for studies not involving humans or animals.

**Informed Consent Statement:** Please add "Informed consent was obtained from all subjects involved in the study." OR "Patient consent was waived due to REASON (please provide a detailed justification)." OR "Not applicable" for studies not involving humans.

**Data Availability Statement:** please refer to suggested Data Availability Statements in section "MDPI Research Data Policies" at https://www.mdpi.com/ethics.

**Acknowledgments:** The authors are grateful to Remy Pires for SEM and EDX analysis and Valérie Lalanne for TEM sample preparation.

**Conflicts of Interest:** The authors declare no conflicts of interest.